\documentclass{article}
\usepackage[T1]{fontenc}
\usepackage[utf8]{inputenc}
\usepackage{tikz}
\usepackage{ismir} 
\usepackage{amsmath,cite,url}
\usepackage{graphicx}
\usepackage{color}
\usepackage{amssymb}
\usepackage[dvipsnames,table]{xcolor}
\usepackage{todonotes}
\usepackage{lipsum}
\usepackage{microtype}    

\usepackage{float} 
\usepackage{pifont} 
\usepackage{url}
\usepackage{multirow}    
\usepackage{makecell}    
\usepackage{rotating}    
\usepackage{booktabs}    
\usepackage{pifont}      
\usepackage{xspace}
\usepackage{array}
\usepackage{arydshln}
\usepackage{float}
\usepackage{caption} 

\newcommand{\thickmidline}[1]{\cmidrule[0.75pt]{#1}}
\usepackage[most]{tcolorbox}

\title{Smorph: Playable Sound Morphing With Diffusion Models}

\multauthor
  {Annie Chu*$^1$ \hspace{0.5cm} Hugo Flores García*$^2$ \hspace{0.5cm} Johannes Imort$^2$ \hspace{0.5cm} \bf Oriol Nieto$^2$}
  {\bf{Bryan Pardo$^1$ \hspace{0.5cm} Jordan Rudess$^3$ \hspace{0.5cm}}
   {\bf Prem Seetharaman$^2$ \hspace{0.5cm} Justin Salamon$^2$}\\
   $^1$ Northwestern University, Evanston, IL, USA\\
   $^2$ Adobe Research, San Francisco, CA, USA\\
   $^3$ Wizdom Music, New York, NY, USA\\
   {\tt\small anniechu@u.northwestern.edu, hflores@adobe.com}\\
    {\href{https://smorphspace.github.io}{\textcolor{blue}{smorphspace.github.io}}}\\
   {\small * denotes equal contribution}
  }
\def\authorname{A. Chu, H. Flores García, J. Imort, O. Nieto, B. Pardo, J. Rudess, P. Seetharaman, and J. Salamon}

\usepackage[bookmarks=false,pdfauthor={\authorname},pdfsubject={\pdfsubject},hidelinks]{hyperref}

\begin{document}

\maketitle
\begin{abstract}\label{sec:abstract}
Sound morphing, generating intermediate sounds that transition from one sonic identity to another, can be a powerful tool for musical sound design. 
Existing diffusion-based morphing approaches entangle temporal structure and timbral identity, offering no mechanism to hold one fixed while transforming the other. 
We present \textbf{\textit{smorph}}, a training-free guidance framework that preserves \textit{how a sound behaves over time} while transforming \textit{what the sound is}, allowing users to morph, for instance from brass to strings at a fixed pitch. We demonstrate across three morphing 
modes: prompt-to-prompt, 
audio-to-prompt, 
and audio-to-audio
.
Evaluations across diverse datasets show that \textit{smorph} effectively produces smooth morph trajectories while substantially improving temporal-structure and source preservation over baselines, albeit with more conservative target-ward transformation in some settings. 
In an exploratory case study, musicians found \textit{smorph} trajectories 
to be expressive and playable, suggesting structural anchoring can serve as a productive constraint for instrumental interaction.

\end{abstract}
\section{Introduction}\label{sec:introduction}
\vspace{-0.5em}
Musicians and sound designers often develop sounds through an iterative process of semantic ideation, audio references, and manual transformation: a producer may envision a \textit{metallic hit}, begin from a gong.wav recording, then modify it to be `sharper' and more `percussive' by mixing it with a \textit{sharp clack}. Sound morphing, i.e. navigating a sequence of intermediate sounds that gradually transition from one sonic endpoint to another, is a natural approach for artists to control and transform sound. Sound morphing can embody a new performative dimension if made interactive, allowing the artist to move fluidly between sonic identities while preserving the gesture or rhythm being played. 

To facilitate musical uses of morphing, it would be desirable to separate high-level audio structure (e.g. rhythms or large-scale pitch contours) from lower-level sonic identity (e.g. timbre) such that each can be controlled independently. If one could combine the ability to specify sonic ideas (either with natural language or audio examples) and interactively morph between them, while preserving high-level structure, the result would be a powerful tool for music performance, sound design, and composition. 
\begin{figure}[t!]
  \centering
  \includegraphics[alt={ISMIR 2025 template example image},width=1\linewidth]{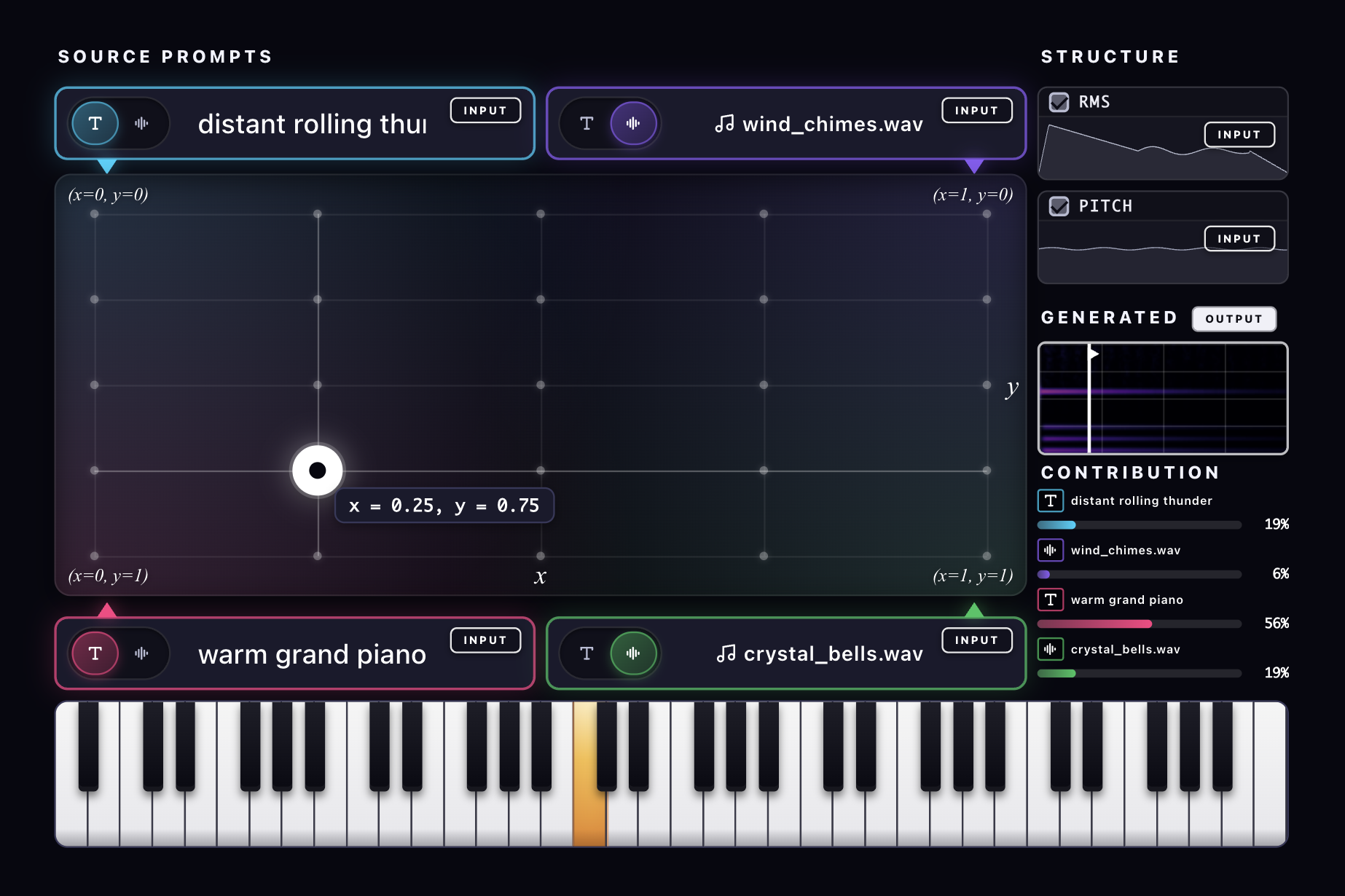}
  \caption{\textbf{\textit{SoundCanvas} by \textit{smorph}}. Four text-or-audio prompts define the corners of a 2D sound morphing space; the $(x,y)$ cursor selects a position in the precomputed $N=25$ sample grid, and real-time bilinear crossfading between neighboring samples provides continuous timbral control. A MIDI keyboard triggers the currently active sound.
}
  \label{fig:overview}
  \vspace{-1.2em}
\end{figure} 
\begin{figure*}[h!]
  \centering
  \includegraphics[alt={...}, width=1.0\linewidth]{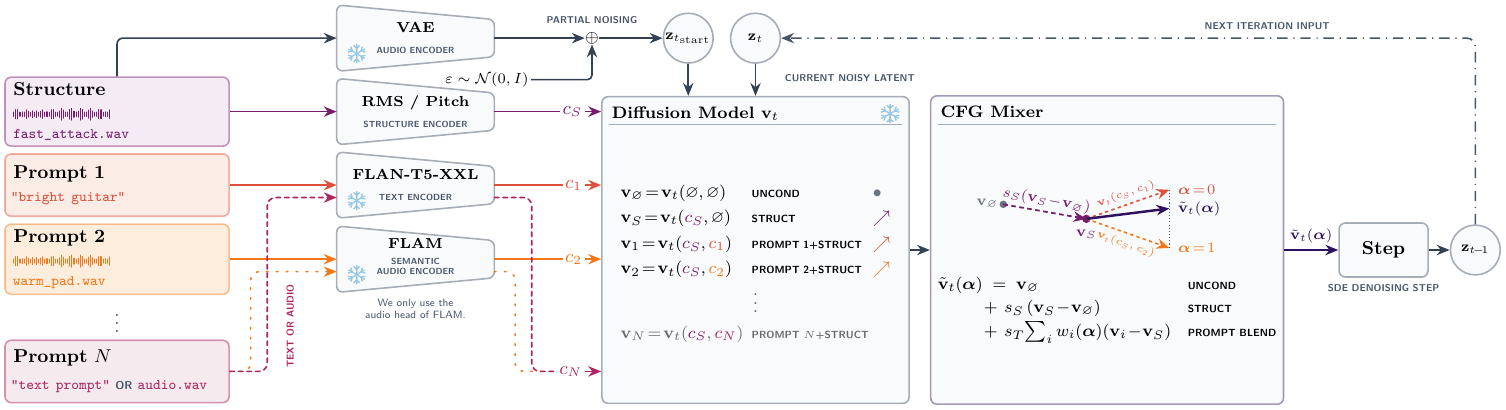}
  \caption{Our structure-guided compositional guidance (See Section \ref{sec:struct-guided-cfg}). The morph parameter $\alpha\in[0,1]$ interpolates audio-/text- endpoint conditions ($c_1$, ...,  $c_N$), while a shared structural reference $c_S$ anchors the temporal envelope or pitch contour across all morph intermediates, fixed across all $T$ denoising steps.}\label{fig:hero}
  \vspace{-0.65em}
\end{figure*} 

While existing approaches achieve portions of this goal, none combine controllable morphing with independent control of structure and timbre. Classical morphing methods based on analysis-synthesis and feature interpolation \cite{caetano2013musical} lack text control and are largely limited to pitched sounds. 
Neural synthesis models (e.g. WaveNet-style \cite{engel2017neural}, RAVE-style VAE models \cite{caillon2021rave}) support broader sound classes, but 
remain restricted to a fixed set of pre-learned sound categories. Related work on style- and timbre-transfer work \cite{verma2018neural, demerle2024style,chu2026mix2morph}, disentangled latent editing \cite{nabi2025adding}, 
and structure-conditioned audio generation systems such as Sketch2Sound \cite{garcia2025sketch2sound} and FlashFoley \cite{novackflashfoley} separate how a sound behaves (\textit{structure}) from what it sounds like (\textit{timbral identity}), but produce a single transferred or generated output rather than a navigable morphing trajectory between endpoints.

More recently, pretrained text-to-audio diffusion models have been used to generate such morphing trajectories, but entangle structure and timbre. MorphFader \cite{kamath2025morphfader} and ComposableArt \cite{valle2025fugatto} interpolate text conditions to support semantically-coherent transitions between prompt-defined endpoints. SoundMorpher \cite{niu2024soundmorpher} interpolates between audio endpoints and re-parameterizes the resulting path via log-Mel distance to improve perceptual spacing. Yet these approaches remain too \textit{permissive}: they vary both temporal structure and timbral identity together. As such, structural elements like rhythm and pitch may drift alongside timbre, and distance-based re-parameterization can only redistribute points along this inherently entangled path.

\textbf{Our work combines text-to-audio generative modeling with sound morphing to produce controllable sound objects for musical interaction.} We treat sound morphing as navigation along a structure-anchored trajectory, using temporal structure as a constraint under which timbral identity changes, and enable morph endpoints to be defined as either text descriptions or audio clips. With these trajectories, we can 
create \textit{semantic controls}--playable timbral axes that allow musicians to transform a sound by sweeping between text- or audio- endpoints in real time. Musicians can thereby intuitively and expressively reshape a sound's character while keeping musically salient cues such as rhythmic timing and pitch contour stable.
We provide demo examples at \textbf{\textcolor{blue}{\href{https://smorphspace.github.io/}{smorphspace.github.io.}}}

\noindent Our contributions are:

    \textbf{C1. Structure-guided sound morphing via compositional guidance.} A classifier-free guidance formulation (Sec. \ref{sec:struct-guided-cfg}) that 
    enables morph trajectories that preserve temporal structure (RMS, pitch contours) while timbral identity transforms continuously between a source and target.
    
    \textbf{C2. Composable multimodal conditioning.} Source and target identities can be specified via combinations of text prompts and audio clips, while high-level structure can be designated via audio-derived controls like RMS and pitch contours (Sec. \ref{sec:struct-guided-morph}, \ref{subsec:editing}, and \ref{subsec:audio2audio}). 
    
    \textbf{C3. Playable Semantic Controls}: Structure-consistent morph trajectories controlled by a playable interface (Fig. \ref{fig:overview}), demonstrated in a case study with musicians (Sec. \ref{sec:smorph-space}).
\vspace{-0.75em}
\section{Background}\label{sec:background}
\vspace{-0.25em}
\subsection{Classifier-Free Guidance (CFG)}\label{subsec:cfg}
\vspace{-0.5em}
Our work generates audio with a diffusion model. Classifier-free guidance (CFG)~\cite{ho2021classifierfree} is the standard mechanism for steering  diffusion towards a conditioning signal. Given a conditioning signal (e.g., a text prompt) $c$, the guided velocity field $\tilde{\mathbf{v}}_t(c)$ is 

\vspace{-0.75em}
\begin{equation}
  \tilde{\mathbf{v}}_t(c)
  = \mathbf{v}_t(\varnothing)
  + \gamma\bigl(\mathbf{v}_t(c) - \mathbf{v}_t(\varnothing)\bigr),
  \label{eq:cfg}
\end{equation}

\noindent where $\mathbf{v}_t(\varnothing)$ is the unconditional denoising prediction, $\mathbf{v}_t(c)$ is the prediction conditioned on~$c$, and $\gamma$ controls the guidance strength. 
\noindent More generally, multi-prompt CFG (e.g.,~\cite{valle2025fugatto}) can be written as:
\vspace{-0.5em}
\begin{equation}
  \tilde{\mathbf{v}}_t
  = \mathbf{v}_\varnothing
  + \gamma \sum_{i=1}^{N} w_i
    \bigl(\mathbf{v}_i - \mathbf{v}_\varnothing\bigr),
  \label{eq:multiprompt}
  \vspace{-0.25em}
\end{equation}

Here, $\mathbf{v}_\varnothing$ serves as a single anchor point and each $\mathbf{v}_i - \mathbf{v}_\varnothing$ is a direction toward prompt~$i$.

InstructPix2Pix~\cite{brooks2023instructpix2pix} introduced a hierarchical extension of CFG for settings where one condition acts as a structural anchor and another specifies the target transformation. 
\begin{equation}
\begin{aligned}
\tilde{\mathbf{v}}_t
&= \mathbf{v}_t(\varnothing,\varnothing) \\
&\quad + s_I\bigl(\mathbf{v}_t(c_I,\varnothing) - \mathbf{v}_t(\varnothing,\varnothing)\bigr) \\
&\quad + s_T\bigl(\mathbf{v}_t(c_I,c_T) - \mathbf{v}_t(c_I,\varnothing)\bigr)
\end{aligned}
\label{eq:ip2p}
\end{equation}

\noindent Here, $c_I$ is the structural condition, $c_T$ is the text condition, and $s_I,s_T$ are independent guidance scales. The key design choice is that the text direction $\mathbf{v}_t(c_I,c_T)-\mathbf{v}_t(c_I,\varnothing)$ is computed relative to the structure-conditioned prediction rather than the unconditional baseline, which allows decoupling of structural preservation from semantic steering. 
\vspace{-1em}
\subsection{SDEdit and Denoising Dynamics}\label{subsec:sdedit}
With a latent audio diffusion model, SDEdit \cite{meng2022sdedit} can edit a \textbf{real} input audio signal by forward-noising its latent representation to a chosen noise level~$t_\text{start}$, then denoising from the resulting~$\mathbf{z}_{t_\text{start}}$ toward a target condition. Different stages of the denoising trajectory determine different aspects of the final sample: coarse structure is established at high noise levels, while texture and detail are refined at lower noise levels~\cite{meng2022sdedit, chu2026mix2morph}. Thus, the choice of~$t_\text{start}$ controls the balance between source preservation vs transformation to target. 

\begin{figure}[h]
  \centering
  \includegraphics[alt={...}, width=1.0\linewidth]{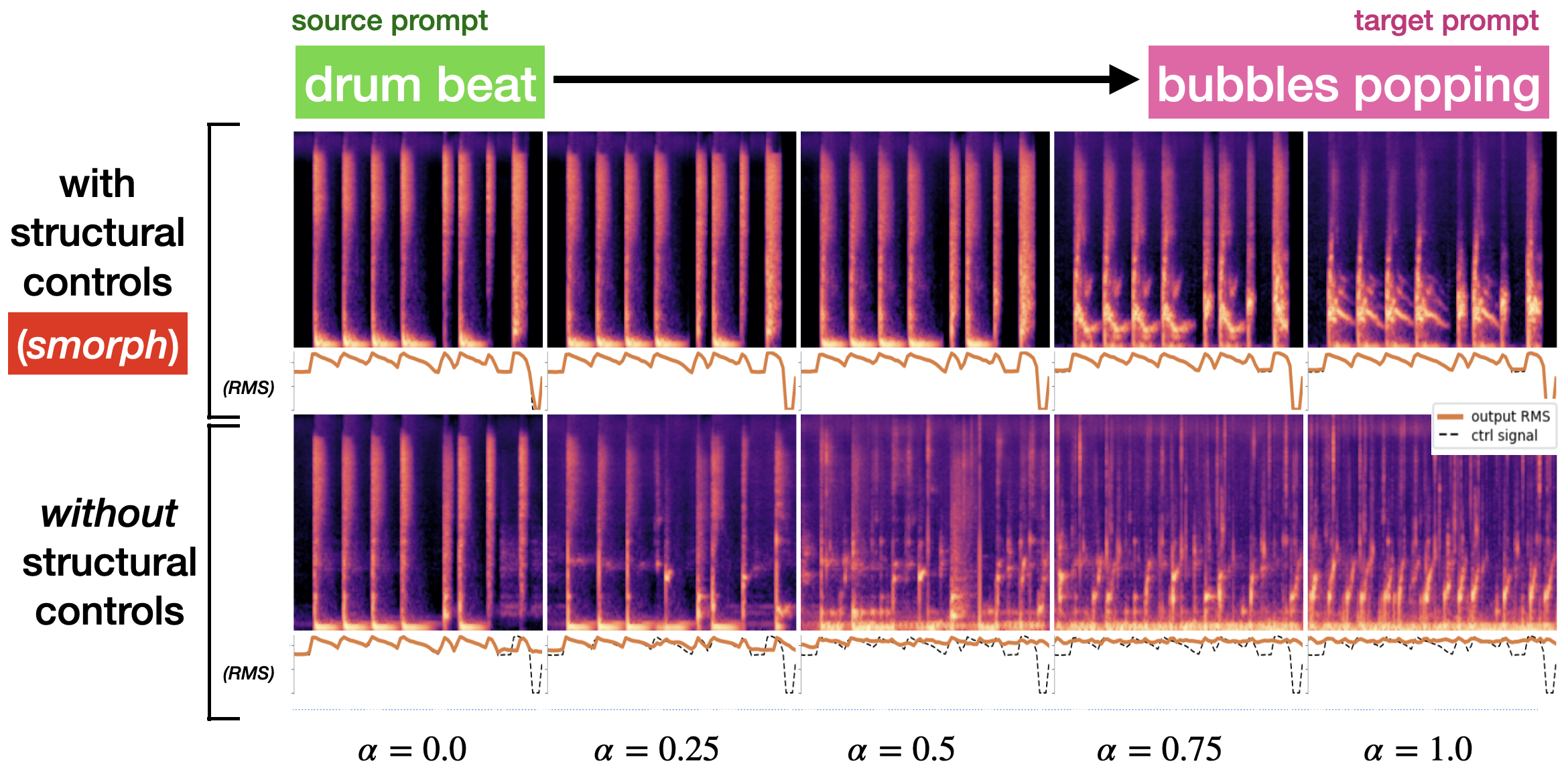}
  \caption{Example morph trajectory with explicit structural control (top) and without (bottom). Each cell shows a mel-spectrogram with input (gray) vs generated (orange) RMS curves.  Explicit control better preserves the intended temporal envelope across the morph trajectory.}
  \label{fig:example_morph}
  \vspace{-1.2em}
\end{figure}
\vspace{-1em}
\section{Method}\label{sec:method}
\vspace{-0.5em}
We introduce a unified compositional guidance formulation for structure-guided sound morphing (C1) with diffusion models. Timbral identity (\textit{what a sound is}) and temporal structure (\textit{how it behaves over time}) are decoupled through composable conditioning signals, enabling multimodal morphing (C2) that preserves desired structure as a source transforms toward a target. Endpoints can be text prompts or audio clips, supporting prompt-to-prompt, audio-to-prompt, and audio-to-audio morphing. In audio-to-X morphing, audio \textit{sources} are initialized via SDEdit\cite{meng2022sdedit}, situating the task closer to semantic editing; audio \textit{targets} serve as timbral \textit{references} rather than waveform content to replicate.

\vspace{-0.5em}
\subsection{Structure-guided compositional guidance (C1)}\label{sec:struct-guided-cfg}
\vspace{-0.5em}
Our framework (Figure \ref{fig:hero}) decomposes guidance into two composable axes — temporal structure via hierarchical two-condition guidance (Eq.\ref{eq:ip2p}), and timbral identity via weighted multi-prompt text composition (Eq.\ref{eq:multiprompt}).

Critically, the timbral term is computed relative to the structure-conditioned prediction $\mathbf{v}_S$ rather than the unconditional baseline $\mathbf{v}_\varnothing$. Given a structural control signal~$c_S$ and $N$ text prompt endpoints $\{c_i\}_{i=1}^N$, we define the guided velocity field as
\vspace{-0.55em}
\begin{equation}
  \tilde{\mathbf{v}}_t(\boldsymbol{\alpha})
  = \mathbf{v}_\varnothing
  + s_{S}\bigl(\mathbf{v}_S - \mathbf{v}_\varnothing\bigr)
  + s_T\sum_{i=1}^{N} w_i(\boldsymbol{\alpha})
    \bigl(\mathbf{v}_i - \mathbf{v}_S\bigr)
  \label{eq:guided}
\end{equation}
\vspace{-1.0em}

\noindent Here $\mathbf{v}_\varnothing = \mathbf{v}_t(\varnothing,\varnothing)$ is the fully unconditional prediction, $\mathbf{v}_S = \mathbf{v}_t(c_S,\varnothing)$ is conditioned on structure only, and $\mathbf{v}_i = \mathbf{v}_t(c_S, c_i)$ is conditioned on \textbf{both} structure and text prompt~$i$. The scalars $s_{S}$ and $s_T$ control structural and text guidance strength respectively. The per-prompt weights $w_i(\boldsymbol{\alpha})$ are determined by morph parameter $\boldsymbol{\alpha}$. For two text prompts ($N=2$), we set \ $w_1\!=\!1{-}\alpha,\;w_2\!=\!\alpha$, where $\alpha\!\in\![0,1]$. For $N>2$, $\boldsymbol{\alpha}$ generalizes to a simplex-valued parameter $[\alpha_1,\ldots,\alpha_N]$, with $\alpha_i\!\geq\!0$ and $\sum_{i=1}^{N}\alpha_i\!=\!1$, setting $w_i(\boldsymbol{\alpha})\!=\!\alpha_i$.
In this case, the structural control signal is provided as specified in Section \ref{sec:struct-guided-morph}.

\noindent \textbf{Computational cost:}
Each denoising step requires $2 + N$ forward passes: one unconditional ($\mathbf{v}_\varnothing$), one control-only ($\mathbf{v}_S$), and one per text prompt ($\mathbf{v}_i$). For the two-prompt morphing case ($N\!=\!2$), this amounts to four passes per step. These forward passes can be executed in a single batch, making the added computation time minimal.
\vspace{-0.5em}
\subsection{Model Details}
\vspace{-0.45em}
We start with a text-to-audio latent diffusion transformer (DiT), previously finetuned to accept time-varying control signals, as in Sketch2Sound \cite{garcia2025sketch2sound}. The base model is a $\sim$1B-parameter DiT 
operating on latents from a KL-regularized DAC-VAE \cite{seetharaman2026taming} that compresses 48kHz mono audio to a 1536-D sequence at 40Hz. Text conditioning uses a frozen \texttt{\small flan-T5-xxl} \cite{chung2024scalingFLAN} encoder via cross attention. This text-to-audio DiT was pretrained on a large mix of proprietary, licensed sound effect datasets and publicly available CC-licensed general audio datasets. 

In our work, time-varying control signals (pitch contour, RMS amplitude) serve as proxies for high-level temporal structure. For pitch extraction, we use CREPE \cite{kim2018crepe, morrison2022torchcrepe}. Signals are injected as additive latent-frame conditioning through dedicated preprocessors. We use k\_diffusion\footnote{github.com/crowsonkb/k-diffusion}'s DPM++2M\-SDE sampling method over a Karras $\sigma$ schedule ($\rho=7.0$, log-SNR endpoints $[-8, 5]$) using $T=24$ denoising steps. Unless otherwise noted, text guidance is $s_T=3.0$ and structural guidance is $s_S=1.0$ (see Eq. \ref{eq:guided}). 

\vspace{-0.5em}
\subsection{Text Prompt-to-Prompt Morphing (C2)}
\vspace{-0.35em}
\label{sec:struct-guided-morph}
We use Eq.~\ref{eq:guided} to perform \textbf{structure-guided sound morphing}. The structural control signal~$c_S$ (e.g.\ an RMS envelope or pitch contour) specifies the temporal behavior that should remain stable across the morph trajectory (Fig. \ref{fig:example_morph}). This signal can be obtained in two ways: self-guided or externally-guided. In the \textbf{self-guided setting}, we can first generate a reference sound using one of the text prompts. 

Similar to Mix2Morph \cite{chu2026mix2morph}, this creates an asymmetrical \cite{chu2026mix2morph} morphing scenario, where one text prompt (i.e. the source prompt) guides the structural shape of the sound, while the target prompt specifies the timbre that the output progressively morphs into. 
In the \textbf{externally-guided setting}, the user provides a reference audio clip from which the structural signal is extracted, allowing the morph to follow a specific temporal envelope, gesture, or pitch contour. In both self and externally guided settings the same guidance equation applies (Eq.~\ref{eq:guided}), only the source of~$c_S$ differs.
\vspace{-0.5em}
\subsection{Source-Preserved Audio-to-Prompt Morphing (C2)}\label{subsec:editing}
\vspace{-0.35em}
While structural control signals (RMS, pitch) anchor higher-level temporal structure, the output may still deviate from the source in more fine-grained aspects (e.g. spectral detail), making it insufficient for small timbral edits to an existing sound (e.g. ``make \underline{this} drum set sound more \textit{metallic}'').
To obtain higher source fidelity at smaller morphing strengths, we augment 
Eq. \ref{eq:guided} with SDEdit\cite{meng2022sdedit}-style initialization, where denoising begins from a partially noised latent rather than pure noise. Given a source audio~$x_\text{src}$ and target text prompt~$c_T$, the source plays two roles:~(1)~its structural descriptors (e.g. RMS, pitch) provide the control signal~$c_S$; (2)~its VAE encoding is forward-diffused to produce ~$\mathbf{z}_{t_\text{start}}$, the initialization point for denoising. Denoising then proceeds from $\mathbf{z}_{t_\text{start}}$ using Eq.~\ref{eq:guided} with $N\!=\!1$, $w_1\!=\!1$. 

As discussed in Sec. \ref{subsec:sdedit}, the noise level~$t_\text{start}$ acts as a morphing parameter in its own right: starting from later steps preserve more of the source's fine detail, while early ``noisier'' steps permit more substantial transformation. Importantly, the structural control signal keeps temporal organization anchored even at high noise levels, effectively extending the usable range of SDEdit beyond what is possible without structural conditioning.
\vspace{-0.5em}
\subsection{Source-Preserved Audio-to-Audio Morphing (C2)}\label{subsec:audio2audio}
\vspace{-0.35em}
When both endpoints are audio recordings $(x_\mathrm{src}, x_\mathrm{tgt})$, we extend our framework via \textit{audio reference conditioning}: target audio is encoded with the FLAM audio encoder~\cite{wu2025flam} and its global embedding replaces the text conditioning vector in Eq.~\ref{eq:guided}. 
To enable this, we use a pre-adapted DiT checkpoint fine-tuned to additionally accept FLAM embeddings (linearly projected to the model's latent dim) as a drop-in text replacement, keeping all CFG formulas unchanged.

This yields two variants. \textbf{FLAM$\to$FLAM} substitutes $\mathrm{FLAM}(x_\mathrm{src})$ and $\mathrm{FLAM}(x_\mathrm{tgt})$ for $(c_1, c_2)$ in Eq.~\ref{eq:guided} and sweeps $\alpha$, with $c_S$ extracted from $x_\mathrm{src}$ -- the audio-reference analogue of prompt-to-prompt morphing. \textbf{SDEdit+FLAM} additionally initializes denoising from the forward-diffused source latent~$\mathbf{z}_{t_\mathrm{start}}$ with $c_1 = \mathrm{FLAM}(x_\mathrm{tgt})$ and sweeps $t_\mathrm{start}$, mirroring the 
audio-to-prompt mode (Sec. \ref{subsec:editing}).
\vspace{-0.5em}
\subsection{Real-Time Morph-Space Interaction}\label{subsec:precompute}
\vspace{-0.35em}
Each point in the morph trajectory requires a full denoising pass through a large (${\sim}$1B parameter) DiT and its latent decoder. Current text-to-sound diffusion models cannot yet stream audio with the sub-100 ms latency~\cite{bartlette2006effect,chew2004musical} required for responsive musical interaction. 
To enable real-time interactive exploration of morphing trajectories, we \textit{precompute} a bank of $N$ samples at uniformly spaced parameter values and store them as a discrete 
bank. For prompt-to-prompt morphing, parameters are uniformly spaced across $\alpha \in [0,1]$; for source-preserving modes (audio-to-X), we select a subset of $t_\mathrm{start}$ values (e.g., indices 5--13 of a 24-step schedule), where higher indices add less noise and remain closer to the source. At interaction time, the user navigates the morph trajectory via slider (or 2D canvas for four-prompt morphing) and receive immediate audio feedback by indexing the bank, with optional equal-gain crossfading between neighboring buffers to allow fractional indexing. The entire bank can be generated in a single batched forward pass: a $N$=25-point, 8s interpolation space completes in under 10 seconds on a single A100 GPU. The bank is recomputed whenever a morph endpoint changes.

\vspace{-1em}
\section{A semantic morph instrument}\label{sec:soundcanvas}
\vspace{-0.35em}
Text-to-audio models generate single, discrete sounds. Inspired by text-to-instrument systems like InstrumentGen~\cite{nercessian2024generating}, we introduce \textbf{\textit{SoundCanvas} (C3)}, which turns a set of text or audio prompts into a sampler instrument with a continuous timbre control that a performer navigates with a 2D crosshair (See Fig. \ref{fig:overview}). 

\noindent \textbf{Construction.} The user assigns four prompts $(c_1,c_2,c_3,c_4)$ to the corners of the canvas 
and optionally provides a structural reference recording $c_S$. We then precompute a $G_x \times G_y$ grid of audio clips under Eq.~\ref{eq:guided}, where each grid position blends the four corner conditions 
via equal-power bilinear interpolation. Clips near a corner sound close to that prompt; clips near the center blend all four. 
\noindent \textbf{Real-time performance.} All $G_x G_y$ precomputed clips are held resident in the audio engine; as the performer moves the crosshair, the four neighboring grid cells are mixed at real-time bilinear, equal-power gains. For pitched instrument patches, we condition the audio grid on a single pitch (e.g. C3), and do live playback resampling to perform different pitches. 
\vspace{-1em}
\section{Objective Experimental Design}
\vspace{-0.25em}
We evaluate whether structure-guided morphing (C1) can preserve musically relevant temporal structure while enabling smooth, expressive timbral transformation. We evaluate via objective experiments on prompt-to-prompt, audio-to-prompt, and audio-to-audio morphing, together with a morph-to-instrument case study (Sec. \ref{sec:smorph-space}) examining whether the resulting morphs support playable musical interaction. 
\vspace{-0.5em}
\subsection{Models Tested}
\vspace{-0.3em}
\noindent \textbf{Our \textit{smorph} variants.} We evaluate two structure-guided variants: \textit{smorph$_e$} is our full explicit-anchor formulation, using the control-only prediction $\mathbf{v}_S$ as a dedicated structural guidance direction (Eq.~\ref{eq:guided}); \textit{smorph$_i$} is an implicit-control ablation that uses the same structural signal and backbone, but does not isolate the control-only prediction $\mathbf{v}_S$ (Eq. \ref{eq:multiprompt}). Unless otherwise noted, \textit{smorph} denotes the explicit-anchor formulation \textit{smorph}$_e$. 
\noindent \textbf{Baselines.} For prompt-to-prompt morphing, we compare against simple audio crossfading (Crossfade) and ComposableART\cite{valle2025fugatto}-style multi-prompt CFG (CoA), which has no explicit high-level structural control. For audio-to-prompt, we compare against plain SDEdit\cite{meng2022sdedit}. 
\textbf{Settings.} All methods share the same denoising backbone; defaults use $s_S=1.0$, $s_T=3.0$.

\vspace{-0.5em}
\subsection{Research Questions}
\vspace{-0.3em}
\textbf{RQ1: How does structural conditioning affect structure-preserving morphing?} 
To answer this, we use prompt-to-prompt morphing on two within-domain settings with different structural demands: Clotho-Clotho, testing RMS preservation across pairs of dynamic SFX and NSynth-NSynth, testing pitch preservation for single-notes. To isolate the effect of structural conditioning, we compare our two structure-guided \textit{smorph} variants (implicit, explicit) against the baselines with no structure guidance. Note that while this mechanism applies to any source--target (X-to-X) morphing, we instantiate via prompt-to-prompt morphing to isolate the effect of structural conditioning from source audio initialization. For each pair $(t_{src}, t_{tgt})$, we generate morph trajectories of $N=21$ samples at $\alpha \in \{0.00, 0.05, \ldots, 1.00\}$ with $w_1=1-\alpha,\; w_2=\alpha$. The structural control signal $c_S$ is extracted from a reference sample generated from $t_{\mathrm{src}}$ (self-guided setting, Sec. ~\ref{sec:struct-guided-morph}).

\textbf{RQ2: How does structural conditioning affect source preservation when morphing from real audio?} 
In the audio-text pair case, $(x_{\mathrm{src}}, t_{\mathrm{tgt}})$, we compare plain SDEdit \cite{meng2022sdedit} against \textit{smorph} (SDEdit + structural control), focusing on source retention across the morph sweep. While structural conditioning applies to any audio-initialized setting, we isolate its effect here before examining the audio-to-audio case in RQ3. For both methods, we sweep $t_{\text{start}}$ over all $T=24$ noise levels, where $t_{\text{start}}$ controls how strongly the real source is transformed towards $t_{tgt}$.

\noindent \textbf{RQ3: Does partial-noise initialization preserve the source better than embedding-only conditioning?} 
In the audio-to-audio case, the source can be introduced either as a latent initialization (SDEdit) or as an embedding-based condition (FLAM). To study this, we evaluate the two variants described in Sec. \ref{subsec:audio2audio}: FLAM$\to$FLAM sweeps $\alpha$ 
($N=21$) using FLAM embeddings of both recordings as conditioning; SDEdit+FLAM sweeps $t_\mathrm{start}$ over all $T=24$ noise levels using the source latent as initialization and $\mathrm{FLAM}(x_\mathrm{tgt})$ as the target condition.

\begin{table}[t]
\centering
\setlength{\tabcolsep}{3pt}
\setlength{\dashlinedash}{1.5pt}
\setlength{\dashlinegap}{1.5pt}
\resizebox{\columnwidth}{!}{%
\begin{tabular}{@{}llccccc@{}}
\toprule
& &

\multicolumn{1}{c}{\textit{Struct.}} &
\multicolumn{2}{c}{\textit{Timbral Trajectory}} &
\multicolumn{1}{c}{\textit{Joint}} &
\multicolumn{1}{c}{\textit{Qual.}} \\
\cmidrule(lr){3-3}
\cmidrule(lr){4-5}
\cmidrule(lr){6-6}
\cmidrule(lr){7-7}
\textbf{Dataset} & \textbf{Method}
    & \makecell{$S_{\mathrm{ctrl}}$ $\uparrow$}
    & \makecell{$\mathrm{Sm}_{\mathrm{tgt}}$ $\uparrow$}
    & \makecell{$\Delta_{\mathrm{FLAM}}$ $\uparrow$}
    & \makecell{\scriptsize{$S_{\mathrm{ctrl}}\!\times\!\mathrm{Sm}_{\mathrm{tgt}}$ $\uparrow$}}
    & \makecell{KAD $\downarrow$} \\
\midrule[0.75pt]

\multirow{4}{*}{\makecell[l]{NSynth-NSynth\\\textit{fixed-pitch notes}\\CTRL=Pitch}}
    & Crossfade
    & 0.048 & 0.324 & \underline{0.155} & 0.016 & 13.602 \\
    & CoA (no ctrl) 
    & 0.012 & \underline{0.442} & 0.154 & 0.005 & 14.058 \\
    & \makecell[l]{\textit{smorph$_i$} (ours)}
    & \textbf{0.140} & \textbf{0.523} & \textbf{0.188} & \textbf{0.073} & \textbf{11.452} \\
    & \makecell[l]{\textit{smorph$_e$} (ours)}
    & \underline{0.094} & 0.307 & 0.130 & \underline{0.029} & \underline{12.587} \\
\midrule[0.75pt]
\multirow{4}{*}{\makecell[l]{Clotho-Clotho\\\textit{dynamic events}\\CTRL=RMS}}
& Crossfade
    & 0.415 & 0.653 & \underline{0.408} & 0.271 & 4.974 \\
    & CoA (no ctrl) 
    & 0.428 & \textbf{0.738} & \textbf{0.415} & 0.316 & 4.641 \\
    & \makecell[l]{\textit{smorph$_i$} (ours)}
    & \textbf{0.771} & 0.616 & 0.316 & \underline{0.475} & \textbf{4.191} \\
    & \makecell[l]{\textit{smorph$_e$} (ours)}
    & \underline{0.733} & \underline{0.712} & 0.359 & \textbf{0.522} & \underline{4.295} \\

\bottomrule
\end{tabular}}
\caption{\textbf{RQ1: Prompt-to-Prompt} morphing with structural control: \textit{\textbf{smorph}} variants outperform CoA \cite{valle2025fugatto}, but effects differ based on dataset settings.}
\vspace{-1.5em}
\label{tab:rq1_results}
\end{table}

\vspace{-0.75em}
\subsection{Objective Evaluation Datasets}
\vspace{-0.45em}

We evaluate using NSynth\cite{engel2017neural}, Clotho\cite{drossos2020clotho}, and FSL-10K\cite{ramires2020} drum recordings. NSynth\cite{engel2017neural} contains pitched, single-note instrument timbres\footnote{we build text captions from NSynth metadata with Claude Opus 4.6}. Clotho\footnote{We apply a text filter to remove clips whose category has predominantly static or slowly varying temporal texture (e.g., \textsc{Ambience}, \textsc{Crowds}, \textsc{Rain}, \textsc{Voices}, \textsc{Air}, and related \textsc{Weather} labels).}\cite{drossos2020clotho} is filtered to emphasize temporally-dynamic everyday sound events with diverse timbral identities, reflecting the kinds of common source material used in sampler-based music making.\footnote{E.g. MF DOOM's ``Doomsday'' samples  \textit{Jeopardy!} ``Daily Double'' SFX; M.I.A's ``Paper Planes'' samples cash registers and gunshot SFX.} 

We construct 100 source-target pairs, first across two within-domain settings: \textbf{NSynth-NSynth}, transforming between instrument timbres at the same fixed \textit{pitch}; \textbf{Clotho-Clotho}, morphing between dynamic sound events while preserving \textit{RMS} envelope. 
For RQ2\&3, we add two additional cross-domain settings: \textbf{NSynth-Clotho}, testing whether a single note's \textit{pitch} persists as notes shift toward a non-musical texture, e.g. violin acquiring the semantic quality of bubbles popping, and \textbf{Drumset-Clotho}, testing whether real FSL-10K \cite{ramires2020} drum samples' RMS dynamics persist as percussive timbres shift toward everyday textures. All samples are 4s; for RQ2\&3, where source audio comes from real recordings, inputs are random 4s excerpts. 
\vspace{-0.55em}
\subsection{Evaluation Metrics}
\vspace{-0.45em}
We evaluate each morph trajectory on structural control adherence, target-directed timbral trajectory, source preservation (RQ2\&3), and audio quality.
\textbf{Structural Adherence.} We extract the structural control signals (RMS, pitch) from generated morphs and compute their correlations with the reference control signal, then average adherence across the morph trajectory. 
\textbf{Timbral Trajectory.} We measure timbral change in FLAM embedding space \cite{wu2025flam}. 
For each morph sample $y_\alpha$, we compute its similarity to the target prompt $\mathrm{sim}_{tgt}(\alpha)$ and report two complementary measures: (a) \textbf{smoothness} (Sm$_{tgt}$), computed as the Spearman correlation between $\alpha$ and $\mathrm{sim}_{tgt}(\alpha)$, measuring whether timbral identity steadily transforms from source-to-target; and (b) \textbf{net transformation} ($\Delta_{\mathrm{FLAM}}$), computed as $\mathrm{sim}_{tgt}(1)-\mathrm{sim}_{tgt}(0)$, measuring the net increase in target similarity from the start vs end of the morph.
\textbf{Joint Adherence-Progression Score (RQ1).} We report $S_{\mathrm{ctrl}} \times \mathrm{Sm}_{\mathrm{tgt}}$ as a compact performance measure of source-structure adherence and semantic progression toward the target.
\textbf{Source Preservation (RQ2\&3).} For audio-to-X morphing, 
we measure source preservation by the L1 distance between the mel-spectrograms of the input and intermediate edited samples at each swept $t_{\text{start}}$ toward the target prompt- (RQ2) or audio- (RQ3) reference.
\textbf{Audio quality.} We report KAD \cite{chung2025kad} over all generated morphs per method w.r.t. reference FLAM \cite{wu2025flam} embeddings computed from a proprietary dataset of 1k high-quality sounds.
\vspace{-0.8em}
\section{Objective Results}\label{sec:results}
\vspace{-0.45em}

\textbf{RQ1 (Table \ref{tab:rq1_results}).} Both \textit{smorph} variants achieve substantially higher structural adherence than CoA across settings, with a \textit{smorph} variant achieving the best joint structure-progression score in each. Comparing the two conditioning regimes: \textit{smorph$_i$} leads on NSynth-NSynth ($0.073$ vs.\ CoA's $0.005$), while \textit{smorph$_e$} leads on Clotho-Clotho ($0.522$ vs.\ CoA's $0.316$), though with a reduced net transformation ($0.359$ vs.\ $0.415$). This suggests explicit structural anchoring is most useful for morphing temporally incongruent sounds, though with some cost to transformation range, whereas implicit joint conditioning can suffice for simpler pitch-preserving morphs where the pitch conditioning might be within-distribution for both source and target endpoints. 
In practice, structural control signals such as RMS envelopes can be strong enough to constrain text guidance~\cite{garcia2025sketch2sound}; accordingly, for explicit structural guidance (\textit{smorph$_e$}), we use a lower structural guidance strength $s_S < s_T$ to allow text semantics to steer the output.

\begin{table}[t]
\centering
\setlength{\tabcolsep}{3pt}
\setlength{\dashlinedash}{1.5pt}
\setlength{\dashlinegap}{1.5pt}
\resizebox{\columnwidth}{!}{%
\begin{tabular}{@{}ll ccccc@{}}
\toprule
& & \multicolumn{2}{c}{\textit{Struct. \& Src. Pres}}
  & \multicolumn{2}{c}{\textit{Timbral Trajectory}}
  & \textit{Qual.} \\
\cmidrule(lr){3-4}\cmidrule(lr){5-6}\cmidrule(lr){7-7}
\textbf{Dataset} & \textbf{Method}
    & \makecell{$S_{\mathrm{ctrl}}$ $\uparrow$}
    & \makecell{mel-L1 $\downarrow$}
    & \makecell{Sm$_{tgt}$ $\uparrow$}
    & \makecell{$\Delta_{\mathrm{FLAM}}$ $\uparrow$}
    & \makecell{KAD $\downarrow$} \\
\midrule[0.75pt]

\multirow{2}{*}{\makecell[l]{NSynth-NSynth\\ CTRL=Pitch}}
    & SDEdit
    & 0.171 & 3.070 & \textbf{0.501} & \textbf{0.227} & 10.452 \\
    & \textit{smorph} (ours)
    & \textbf{0.225} & \textbf{2.623} & 0.500 & 0.198 & \textbf{10.279} \\

\thickmidline{1-7}

\multirow{2}{*}{\makecell[l]{NSynth-Clotho\\ CTRL = Pitch}}
    & SDEdit
    & 0.144 & 3.542 & \textbf{0.741} & \textbf{0.540} & \textbf{4.158} \\
    & \textit{smorph} (ours)
    & \textbf{0.187} & \textbf{3.053} & 0.673 & 0.466 & 5.579 \\

\thickmidline{1-7}

\multirow{2}{*}{\makecell[l]{Clotho-Clotho\\CTRL=RMS}}
    & SDEdit
    & 0.356 & 1.518 & \textbf{0.695} & \textbf{0.392} & \textbf{4.420} \\
    & \textit{smorph} (ours)
    & \textbf{0.646} & \textbf{1.027} & 0.646 & 0.356 & 4.902 \\

\thickmidline{1-7}

\multirow{2}{*}{\makecell[l]{Drumset-Clotho\\CTRL=RMS}}
    & SDEdit
    & 0.443 & 2.115 & \textbf{0.807} & \textbf{0.559} & \textbf{5.512} \\
    & \textit{smorph} (ours)
    & \textbf{0.732} & \textbf{1.418} & 0.743 & 0.441 & 6.822 \\

\bottomrule
\end{tabular}}
\caption{\textbf{RQ2: Audio-to-Prompt morphing.} Structural control improves source preservation over SDEdit \cite{meng2022sdedit}, with a tradeoff in target transformation and quality.}
\vspace{-1.5em}
\label{tab:rq2_results}
\end{table}

\noindent\textbf{RQ2 (Table \ref{tab:rq2_results}).}
Adding structural control to SDEdit consistently improves source preservation (mel-L1): \textit{smorph}~(SDEdit+ctrl) achieves lower mel-L1 than plain SDEdit on all datasets, confirming that structural guidance anchors generation to the source audio. This stronger anchoring is accompanied by a target reach and audio-quality cost, reflected by lower net transformation ($\Delta_{\mathrm{FLAM}}$) on all four settings and SDEdit's lower KAD on three of the four. These results suggest that structural guidance constrains timbral drift from the source, limiting how freely the diffusion process can optimize for perceptual quality.

\textbf{RQ3 (Table \ref{tab:rq3_results}).}
SDEdit+FLAM improves source preservation (lower mel-L1) and audio quality (KAD) over FLAM$\to$FLAM across all settings, suggesting that source-latent initialization retains source information without an apparent quality tradeoff observed with stronger structural conditioning in RQ2. Structural adherence varies by control type: SDEdit+FLAM leads on pitch-conditioned settings and FLAM$\to$FLAM leads on RMS, while target-directed trajectory metrics vary by dataset.


\vspace{-1.20em}
\section{Subjective User Evaluation}\label{sec:smorph-space}
\vspace{-0.45em}
To explore how musicians might use \textit{SoundCanvas} (Sec. \ref{sec:soundcanvas}), we conducted an exploratory user study with practicing musicians ($N=6$; ages 26-37; 2-28 years of production experience). 
In a 55-minute session, each participant explored the 2D text-defined morph canvas connected to a pitch-mapped keyboard (Fig. \ref{fig:overview}); endpoints limited to text-only. Participants constructed morphspaces manually by designating all four endpoints, 
and then experimented with varying control signals: RMS, Pitch (on/off). Sessions concluded with a semi-structured interview focused on musical utility, playability, and expressive reach. 
 This was followed by a post-session Creativity Support Index (CSI) survey \cite{cherry2014quantifying}.

From session recordings, we identified two recurring themes using a reflexive thematic analysis \cite{braun2006using, braun2019reflecting}: making uncertainty playable through structured serendipity (T1) and enabling continuous sonic palettes from discrete parameters (T2). We report unweighted mean item ratings ($M$) from the post-session CSI survey.

\textbf{T1. ``Structured'' Serendipity: uncertainty made playable.} 
\textit{SoundCanvas} made unfamiliar material playable through familiar interaction paradigms, similar to the interfaces found in musical instruments like the Portal (P2, P3) and  XY/Kaoss-style pads (P4). What made it compelling was that traversal through the space enabled discovery of novel material that made musical sense. 

Sounds produced by movement through the 2D space felt predictable in direction and smooth in traversal due to structural anchoring: quadrants \textit{``guide you to where you’re going''} and participants \textit{``knew exactly where [they] needed to go''} to reach particular timbres (P3), with \textit{``no point where the scaling of the movement felt off''} (P2). Traversal was broadly perceived as  ``satisfying'' - \textit{``every point had some value to add sonically''} (P2) - though some semantically dissonant prompt combinations produced uneven ``hot spots'' (P4). This was consistent with high CSI ratings for \textit{exploration} ($M=8.3/10$), \textit{enjoyment} ($M=8.5$), \textit{satisfaction} ($M=8.5$), and \textit{results-worth-effort} ($M=9.0$).

Crucially, 
intermediate sounds in the canvas were not merely transitional between referents, but legible musical objects in and of themselves.  
Interior regions were sites of \textit{``valuable surprises''} (P2) where finding \textit{``happy accidents''} (P3) was \textit{``kind of inevitable''} (P2); participants wanted to export and reuse these points (P1, P3, P4, P6). 
Together, these responses suggest \textit{SoundCanvas} was playable not because it removed uncertainty, but because it made uncertainty navigable.
 This is reflected in CSI scores for \textit{creativity} ($M=8.0/10$) and expressiveness ($M=7.8/10$).

\begin{table}[t]
\centering
\setlength{\tabcolsep}{3pt}
\setlength{\dashlinedash}{1.5pt}
\setlength{\dashlinegap}{1.5pt}
\resizebox{\columnwidth}{!}{%
\begin{tabular}{@{}ll ccccc@{}}
\toprule
& & \multicolumn{2}{c}{\textit{Struct. \& Src. Pres}}
  & \multicolumn{2}{c}{\textit{Timbral Trajectory}}
  & \textit{Qual.} \\
\cmidrule(lr){3-4}\cmidrule(lr){5-6}\cmidrule(lr){7-7}
\textbf{Dataset} & \textbf{Method}
    & \makecell{$S_{\mathrm{ctrl}}$ $\uparrow$}
    & \makecell{mel-L1 $\downarrow$}
    & \makecell{Sm$_{tgt}$ $\uparrow$}
    & \makecell{$\Delta_{\mathrm{FLAM}}$ $\uparrow$}
    & \makecell{KAD $\downarrow$} \\
\midrule[0.75pt]

\multirow{2}{*}{\makecell[l]{NSynth-NSynth\\ CTRL=Pitch}}
    & FLAM$\to$FLAM
    & 0.120 & 3.766 & 0.112 & 0.056 & 11.095 \\
    & SDEdit+FLAM
    & \textbf{0.204} & \textbf{2.772} & \textbf{0.150} & \textbf{0.080} & \textbf{10.173} \\

\thickmidline{1-7}

\multirow{2}{*}{\makecell[l]{NSynth-Clotho\\ CTRL=Pitch}}
    & FLAM$\to$FLAM
    & 0.057 & 4.722 & \textbf{0.623} & \textbf{0.287} & 4.520 \\
    & SDEdit+FLAM
    & \textbf{0.159} & \textbf{3.429} & 0.589 & 0.271 & \textbf{4.485} \\

\thickmidline{1-7}

\multirow{2}{*}{\makecell[l]{Clotho-Clotho\\ CTRL=RMS}}
    & FLAM$\to$FLAM
    & \textbf{0.666} & 1.177 & \textbf{0.489} & \textbf{0.256} & 6.092 \\
    & SDEdit+FLAM
    & 0.650 & \textbf{1.117} & 0.445 & 0.249 & \textbf{5.396} \\

\thickmidline{1-7}

\multirow{2}{*}{\makecell[l]{Drumset-Clotho\\ CTRL=RMS}}
    & FLAM$\to$FLAM
    & \textbf{0.714} & 2.066 & 0.725 & 0.383 & 8.104 \\
    & SDEdit+FLAM
    & 0.693 & \textbf{1.498} & \textbf{0.731} & \textbf{0.415} & \textbf{6.889} \\

\bottomrule
\end{tabular}}
\caption{\textbf{RQ3: Audio-to-Audio morphing.} Partial denoising improves preservation over FLAM-only conditioning, while trajectory \& structure metrics vary by setting.}
\vspace{-1.5em}
\label{tab:rq3_results}
\end{table}

\textbf{T2. Parameters to palettes}. 
Instead of editing low-level parameters or selecting from fixed sample libraries, \textit{SoundCanvas} afforded users movement through continuous sonic palettes of related timbres and affective states. 

Participants described varying the use of structural conditioning (pitch, RMS) at different moments in their creative practice. Turning off structure controls was valued for ideation (P4, P6), \textit{``maximizing [timbral] variety''} (P1) and creating \textit{``crazy textures''} (P6).  Participants valued RMS-on as a middle ground: a temporally-grounded, yet timbrally-rich exploration space suited for percussion-oriented workflows (P2), groove realization (P1), and rhythmic ideas (P5). Pitch-on, in contrast, was described more explicitly as performable: it felt \textit{``cohesive...like [an] expressive gesture''} (P5), making the canvas feel more traditionally instrument-like and performance-ready (P5, P6). 
Participants also described \textit{SoundCanvas} as a way to expand and recontextualize existing sonic assets: \textit{``exploding''} personal catalogs into navigable palettes (P3), making recognizable popular sample packs more unique (P6), or salvaging forgotten recordings into usable material (P4). 

\vspace{-0.75em}
\section{Conclusions}
\vspace{-0.35em}
This paper explored generative sound morphing as a way to provide controllable musical objects. 
A musician may seek to transform timbre while preserving musically-relevant structural features (e.g rhythm or pitch contours). 
We offer a training-free CFG formulation with an explicit structural term, better balancing control adherence with timbral progression, 
with musician feedback highlighting structural constraints as central to trajectory playability. However, structure-preserving morphing remains an open challenge. 
With multiple conditioning signals, one can dominate the other, and a trajectory that appears smooth in embedding space may not feel perceptually linear or musically meaningful. Future work may integrate perceptual linearization improvements with richer interaction mechanisms to make morphs more navigable, expressive, and performable.



\newpage
\vspace{-1.0em}
\section{Acknowledgments}
\vspace{-0.5em}
The authors would like to thank study participants M. Jones, M. McCormack, R. Gao, J. Thompson, R. Devens, M. Burton for their time, feedback, and generous participation in testing \textit{SoundCanvas}.
\vspace{-1.0em}
\section{AI Usage Statement}
\vspace{-0.5em}
The authors used Claude (an LLM) for minor text editing, typo parsing, and cleanup, as well as drafting figures (via TikZ code). The technical ideas, experimental design, analysis, claims, and final writing decisions are all our own. The authors used Claude code to develop the demo interface and web UI. No experimental data, results, or citations were generated by AI tools.
\vspace{-1.0em}
\section{Ethics Statement}
\vspace{-0.35em}
Our user study involved a small group of expert musicians who provided informed consent before participation. Participants interacted with generated audio examples and were invited to continue accessing the demo after the study. We recognize that generative audio systems may reflect biases in their training data and conditioning interfaces, including bias toward English-language prompts and particular Western production practices. We also recognize broader concerns around authorship, attribution, and the reuse or transformation of existing sounds. While this work hopes to provide creators with new timbral possibilities, and to contribute to a long history of recorded sound as material for composition, production, and performance (e.g., sampling, remixing, turntablism, etc), we recognize that systems like this should be developed with continued attention to access, attribution, and the ways generated sounds enter existing creative economies.
\bibliography{references}

\end{document}